\documentstyle[12pt]{article}

\def\thefootnote{\fnsymbol{footnote}}

\newcommand{\eq}{\begin{equation}}
\newcommand{\en}{\end{equation}}
\newcommand{\eqa}{\begin{eqnarray}}
\newcommand{\ena}{\end{eqnarray}}

\newcommand{\br}{\langle}
\newcommand{\kt}{\rangle}
\newcommand{\um}{\frac12}

\newcommand{\NP}[1]{Nucl.\ Phys.\ {\bf #1}}
\newcommand{\PL}[1]{Phys.\ Lett.\ {\bf #1}}

\newcommand{\PR}[1]{Phys.\ Rev.\ {\bf #1}}
\newcommand{\PRL}[1]{Phys.\ Rev.\ Lett.\ {\bf #1}}

\begin{document}
\begin{titlepage}
\vskip0.5cm
\begin{flushright}
DFTT 68/95\\
DAMTP 95-60 \\
\end{flushright}
\vskip0.5cm
\begin{center}
{\Large\bf Deconfinement  Transition and Dimensional}
\vskip0.2cm
{\Large\bf  Cross-over in the 3D Gauge Ising Model}
\end{center}
\vskip 1.3cm
\centerline{M. Caselle$^a$ and M. Hasenbusch$^b$ }
\vskip 1.0cm
\centerline{\sl  $^a$ Dipartimento di Fisica 
Teorica dell'Universit\`a di Torino}
\centerline{\sl Istituto Nazionale di Fisica Nucleare, Sezione di Torino}
\centerline{\sl via P.Giuria 1, I-10125 Torino, Italy
\footnote{e--mail: caselle~@to.infn.it}}
\vskip .4 cm
\centerline{\sl $^b$ DAMTP, 
Silver Street, Cambridge, CB3 9EW, England
\footnote{e--mail: M.Hasenbusch@damtp.cam.ac.uk}}
\vskip 1.cm

\begin{abstract}
We present a high precision Monte Carlo study of the finite temperature 
$Z_2$ gauge theory in $2+1$ dimensions. The duality with the 3D 
Ising  spin model allows us to use powerful cluster algorithms for the
simulations.  
For temporal extensions of up to $N_t=16$ we obtain the inverse critical 
temperature with a statistical accuracy comparable with the most accurate 
results for the bulk phase transition of the 3D Ising model.  
We discuss the predictions of T. W. Capehart and M.E. Fisher  for the
dimensional crossover from 2 to 3 dimensions. 
Our precise data for the critical exponents and critical amplitudes confirm the 
Svetitsky-Yaffe conjecture. We find deviations from Olesen's 
prediction for the critical temperature of about 20 $\%$. 
\vskip0.2cm
\end{abstract}
\end{titlepage}

\setcounter{footnote}{0}
\def\thefootnote{\arabic{footnote}}

\section{Introduction}
One of the most interesting predictions of QCD  is the existence of a 
deconfinement transition at some critical temperature $T_c$.
Finding a precise characterization of this transition is, however, still 
an open problem. In particular it would be interesting to have some 
indication on the order of the transition and, in case it is of second 
order, on its critical indices.
The natural framework to pose these questions is that of the finite 
temperature Lattice Gauge Theories (LGT). In this framework, a seminal 
contribution was given more than ten years ago by B. Svetitsky and 
L.Yaffe~\cite{sy}. They showed that, if the deconfining transition of a 
given $(d+1)$ dimensional gauge theory is of 
second order, then its universality class should coincide with that of 
the $d$-dimensional spin model, with the center of the original gauge group 
as symmetry group. This result is usually known as the 
``Svetitsky-Yaffe (SY) conjecture'' and has been confirmed in these last 
ten years by several Montecarlo simulations. Most of these tests were made on
LGT with continuous gauge groups (in particular SU(2) and SU(3)) which are
obviously the most interesting from a phenomenological point of view, but do
not allow in general to reach the high statistics and precision which is needed
to test the critical behaviour of a statistical system near a second order
phase transition. To make such a test we follow the opposite
route, namely to choose the simplest possible nontrivial LGT, 
 and then try to reach in this simplified model the highest possible
confidence. Therefore we shall concentrate in the present
paper on the three dimensional gauge Ising model.
Besides its simplicity there is another important reason to choose exactly the
Ising model to test the SY conjecture. 
The only claim of a failure of 
the SY conjecture~\cite{wz} that we are aware of 
  was indeed obtained in a high precision
study of the 3d gauge Ising model with temporal extent $N_t=8$. 
Such a result, if confirmed, would obviously imply 
a deep change in our understanding of the deconfinement transition. In 
this paper we shall address this question, pushing the analysis up to 
$N_t=16$, again in the 3d Ising gauge model, using a new original 
algorithm and very high statistics. We shall show that all our results 
are in perfect and complete agreement with the SY conjecture, we shall 
also give some argument to justify the apparent disagreement found 
in~\cite{wz}.
In order to test the SY conjecture we need, first of all, a very precise 
estimates of the critical deconfinement temperature. With this estimate 
we can test another interesting conjecture, due to Olesen~\cite{ol}, 
which is completely independent from the SY one, and is based on the 
string (or flux tube) description of the infrared regime of QCD. 

\vskip 0.2cm

A seemingly independent, interesting problem, which is actually related to the 
above one, is that of the so called ``dimensional crossover'' in layered 
spin systems. Indeed it is well known that the dimensionality of a 
system as it approaches the critical point (and hence the universality 
class to which the system belongs) is only determined by the number of 
spatial dimensions in which the system has an infinite extension. For 
instance, the Ising model on a lattice with a finite thickness $N_t$ in 
one direction (and the lattice size in the other two directions, say 
$N_s$, going to infinity) belongs to the same universality class of the 
ordinary two-dimensional Ising model. As $N_t$ increases
 both the critical indices of 
the two and of the three-dimensional Ising model can be observed. In 
particular, it is possible to define so called ``effective critical 
indices'' which smoothly interpolate between the $d=2$ to the $d=3$ 
value of the indices as the adimensional ratio $N_t/\xi(\beta)$ (where 
$\xi(\beta)$ is the bulk correlation length of the system) goes from 
zero to infinity. The theory of dimensional crossover was developed 
 by T. W. Capehart and M. E. Fisher~\cite{cf} as a 
particular case of the more general theory of finite size scaling 
theory. Their predictions were then compared with high temperature 
expansions finding in general a good agreement. It is interesting to 
notice that even if ref.~\cite{cf} and ref.~\cite{sy} use different 
languages and are interested in different systems they follow actually 
the same type of arguments (scaling and renormalization group) and reach 
in fact exactly  the same conclusions. This is particularly evident in 
the case of  the three-dimensional Ising model, where dimensional 
cross-over and deconfinement transition are in fact the two faces of 
the same phenomenon. Duality, which maps the Ising gauge model onto the 
spin model also relates  deconfinement and dimensional cross-over.
There is however in~\cite{cf} a very puzzling result. The critical index 
$\lambda$
with which the critical temperature on lattices of finite thickness 
approaches the bulk critical temperature turns out to be different for 
different boundary conditions. In particular, in the case of periodic 
boundary conditions in which we are interested here, 
it does not coincide with 
the value $\lambda=1/\nu$ (where $\nu\sim0.63$,
 is the critical index of the correlation length)
 which one would expect. This is again a
result which, if confirmed, would change our approach to the analysis 
of the deconfinement transition, since it is a common and general use to
extrapolate the finite 
lattice determinations of the critical temperature to the continuum 
limit, following the scaling behaviour of the correlation length.
 In this paper we shall also address this question, showing that
the correct answer is indeed $\lambda=1/\nu$ and that the result 
of~\cite{cf} is probably an artifact of the strong coupling expansion.

\vskip 0.2cm

A further reason of interest in the 3d Ising model on layered lattices is that
in these last years this model (and in particular its realization on lattices
with very few layers) has become a laboratory to test various theoretical
methods~\cite{ls,acpv,mhz}
 to find the critical temperature and the critical indices. Due to the
high precision of our results we are in the position to test the reliability
of these methods and, at least in one case (the MHZ approach~\cite{mhz}, see
below) to better understand the reason of its failure.

\vskip 0.2cm

This paper is organized as follows. Sect.2  will be devoted to 
a brief introduction to finite temperature LGT and to the 
Svetitsky-Yaffe conjecture. In sect.3 we discuss dimensional crossover 
following~\cite{cf}, while in sect.4 a brief account of the string model 
approach to the infrared regime of LGT, and the consequent Olesen 
conjecture for the deconfinement temperature will be given.
In sect.5 we shall describe the model and the  algorithm that we use.
Finally section 6 will be devoted to a discussion of the results of the 
simulation and sect. 7 to the comparison with other existing estimates of the
critical temperature and of the critical indices.

\section{Finite Temperature LGT}

\subsection{General setting}

Let us consider a pure gauge theory with gauge group $G$, defined
on a $d+1$ dimensional cubic lattice.
In order to describe a finite temperature LGT,
we have to impose periodic
boundary conditions in one direction (which we shall call from now
on ``time-like'' direction), while the boundary conditions in the
other $d$ directions (which we shall call ``space-like'') can be
chosen freely.
We take a lattice of $N_t$ ($N_s$) spacings in the time (space)
direction, and we work with the pure gauge theory, containing
only gauge fields described by the link
variables $g_{n;\mu} \in G$, where $ n \equiv (\vec x,t)$
denotes the space-time position of the link and $\mu$ its direction.
We shall denote the time-like direction with the index $\mu=0$ and the 
remaining space-like directions with roman indices $\mu=i=1\cdots d$.
We choose for simplicity the same bare coupling $\tilde\beta$ in the time and 
space directions. The Wilson action is then
\eq
S_W=\sum_{n}~\left\{\tilde\beta\left(\sum_i~{\rm Tr_f}(g_{n;0i})
+\sum_{i<j}~{\rm Tr_f}(g_{n;ij})\right)\right\}~~~,
\label{wilson}
\en
where ${\rm Tr_f}$ denotes the trace in the fundamental representation
and $g_{n;0i}$ ($g_{n;ij}$) are the time-like (space-like)
plaquette variables, defined as usual by
\eq
g_{n;\mu\nu}=g_{n;\mu}g_{n + \mu;\nu}
g^\dagger_{n + \nu;\mu}g^\dagger_{n;\nu}~~~.
\label{plaq}
\en

In this framework the temperature $T$ is given by:
\eq
T=\frac{1}{N_ta}~~~,
\label{temperature}
\en
where $a$ is the  lattice spacing.

In a finite temperature discretization it is possible to define
gauge invariant observables which are topologically non-trivial,
as a consequence of the periodic boundary conditions in the
time direction.
The simplest choice is the Polyakov loop,
defined in terms of link
variables as
\eq
P(\vec x)\equiv
{\rm Tr_f}~V(\vec x)= {\rm Tr_f} \prod_{t=1}^{N_t}(g_{\vec x,t;0})~~~.
\label{polya}
\en

A second important consequence of the periodic boundary conditions in 
the time direction is that the theory  has a new
global symmetry (unrelated to the gauge symmetry), with the center $C$ of the 
original gauge group as symmetry group (hence, if $G$ is abelian, $C$ will 
coincide with $G$). The Polyakov loop
is a natural order parameter for this symmetry.

In $d>1$, finite temperature gauge theories admit a
deconfinement transition at $T=T_c$, separating the
high temperature, deconfined, phase ($T>T_c$) from the low
temperature, confining domain ($T<T_c$). The high
temperature regime is characterized by the  breaking of the global
symmetry with respect to the center of the group. In this phase the
Polyakov loop has a non-zero expectation value, and it is
an element of the center of the gauge group.

\subsection{Svetitsky-Yaffe conjecture}

The idea on which the SY conjecture is based is that, if one would be 
able to integrate out all the 
gauge degrees of freedom of the original $(d+1)$--dimensional
 model {\sl except those related to the Polyakov loops}
then the resulting 
effective theory for the Polyakov loops would be a $d$-dimensional spin 
system with symmetry group $C$. The deconfinement transition of the 
original model would become the order--disorder transition of the 
effective spin system.
This effective theory would obviously have very complicate 
interactions, but Svetitsky and Yaffe were able to argue that all these 
interactions should be short ranged. As a consequence, if the 
transition point of the
effective spin system is of second order, near this critical point, 
where the correlation length becomes infinite, the precise form of 
the short ranged 
interactions should not be important, and the universality class of the 
deconfinement transition should coincide with that of the simple spin 
model with only nearest neighbour 
interactions and the same global symmetry group. 
As a consequence all the critical indices describing the two 
transitions  and all the adimensional ratios of 
scaling quantities should 
coincide in the limit. This is the case for instance of the ratio of 
partition functions which will be discussed in sect.5 and of the 
indices $\nu$ and $\eta$ which will be discussed in sect.6.

This argument holds also for abelian gauge groups. In particular, 
the deconfinement transition of the (2+1) dimensional gauge Ising model 
which we shall study in the following should belong to the same 
universality class as the two-dimensional spin Ising model.

\subsection{Duality}

A peculiar, interesting, feature of {\sl abelian}
 LGT in 2+1 dimensions is 
that they can be mapped through duality, into (2+1) dimensional spin 
models, with the same symmetry group (for a review see for 
instance~\cite{sav}). 

In the case of the Ising model (in which $g_{n,\mu}\in \{\pm 1\}$)
 this is accomplished by the well known
Kramers--Wannier duality transformation which relates the two partition 
functions:
\eqa
Z_{gauge}(\tilde\beta)~\propto~ Z_{spin}(\beta)&&\\
{\beta}=-\um\log\left[\tanh(\tilde\beta)\right]~~&&~~,
\label{dual}
\ena
where $Z_{gauge}(\tilde\beta)$ is defined, according to eq.(\ref{wilson}) and
(\ref{plaq}) as
\eq
Z_{gauge}(\tilde\beta)=\sum_{\{g_{n,\mu}=\pm1\}}\exp\left(\sum_{n,\mu\nu}
\tilde\beta g_{n;\mu\nu}\right)~~~,
\en
and
\eq
Z_{spin}({\beta})=\sum_{s_i=\pm1}\exp(-\beta H(s))~~~,
\label{zspin}
\en
with
\eq
H(s)=-\sum_{\br ij \kt}J_{\br ij \kt}s_is_j
\label{hspin}
\en
where $s_i\in\{\pm1\}$, $i$ and $j$ denote nodes of the dual lattice and the sum is 
extended to the links ${\br ij \kt}$ connecting  the nearest--neighbour 
sites. For the moment the couplings $J_{\br ij \kt}$ are all chosen 
equal to $+1$ . 

Using the duality transformation it is possible to build 
up a one--to--one mapping of physical observables of the gauge system 
onto the corresponding spin quantities. In particular the deconfinement 
transition in the gauge Ising model becomes the order-disorder 
transition in the spin Ising model.

It is important to notice that in general the duality 
transformation does not conserve specific choices of
the boundary conditions (bc in the following). 
In fact, in obtaining eq.(\ref{dual}) (see for 
instance~\cite{sav}) one usually assumes to be in the thermodynamic 
limit and neglects bc. 
In our discussion it is crucial to impose 
{\sl periodic} bc in the time direction in the gauge 
model, but there is no reason to expect that these hold also for the dual 
spin model.
In fact, it can be shown that the corresponding dual model is then
a spin Ising model with 
{\sl fluctuating} bc in the time direction, namely a 
mixture of the two partition functions with periodic and antiperiodic 
bc respectively. However the antiperiodic bc in the time direction 
force the formation of a space-like interface in the Ising spin model. 
As a consequence the antiperiodic free energy will be depressed 
with respect to the periodic one by a factor proportional to $\exp(-
\sigma N_xN_y)$ and in the thermodynamic limit $N_x,N_y \to \infty$
it will become negligible.
 Hence in 
this limit the duality transformation actually becomes a 
mapping between a spin and a gauge Ising model, {\sl  both} with 
periodic bc in the time direction. This will allow us to make the 
simulations directly in the {\sl spin} Ising model with periodic bc 
(which is a very 
convenient choice, since for this model (see sect.5) we have a very 
powerful algorithm to extract the critical couplings) and then to 
translate the result in the framework of the {\sl gauge} Ising model, 
so as to test the SY and Olesen conjectures.

Since this equivalence is of central importance in our work it could 
be useful to discuss it from a different (but equivalent) point of view. 
We can immediately see that a mapping between gauge and spin model both 
with periodic bc is impossible because there is no dual counterpart 
for the expectation value of an isolated Polyakov loop. Since the 
Polyakov loop is the order parameter of the deconfinement transition 
this seems to exclude the possibility to study this phase transition in 
the dual model. However the dual 
of the correlator between two Polyakov loops is perfectly well defined. 
Given any surface in the gauge model bordered by the two Polyakov loops
the dual of the correlator  is given by $Z_f/Z_{spin}$ where $Z_f$
is the partition function of the spin model in which the links 
orthogonal to the surface have been frustrated. If we now take the 
thermodynamic limit in the spacelike directions (and only in this limit)
we can push away as far as we want the two Polyakov loops and obtain, 
by using the standard factorization theorems, the expectation value 
of the isolated Polyakov loop.

\section{Dimensional cross-over in spin models}

In the study of dimensional cross-over in spin models the $N_t$ 
space-like planes are usually called layers.
Let us assume that the pure $d$-dimensional model (one layer) has a 
critical point located at $\beta=\beta_c(N_t=1)$. 
If the $d$ spatial dimensions are sent to infinity, keeping the number of 
layers finite, the system still has a critical point, whose position 
$\beta=\beta_c(N_t)$ changes continuously with $N_t$. The limit
$\lim_{N_t\to \infty} \beta_c(N_t)$ will coincide with the (bulk) 
critical point (which we shall simply denote as $\beta_c$ in the 
following) of the $(d+1)$ model. The reduced critical point shift is 
defined by:
\eq
\epsilon(N_t)=\frac{\beta_c-\beta_c(N_t)}{\beta_c}
\label{3.1}
\en
We can introduce the critical point shift exponent as
\eq
\epsilon(N_t)\sim b_1 \; N_t^{-\lambda} \hskip 1cm  N_t \to \infty
\label{3.2}
\en
with $b_1$ a suitable constant.
Let us introduce the reduced temperature (with respect to the bulk transition 
of the $d+1$ model) :
\eq
t=\frac{\beta_c-\beta}{\beta_c}
\en
Then the bulk correlation length has the following scaling behaviour:
\eq
\xi(\beta) \sim b_2 \; t^{-\nu}  \hskip 1cm  N_t \to \infty
\en

If we assume that the critical point in the layered system is reached 
when $\xi(\beta_c(N_t))\sim N_t$ then it immediately follows that
$\lambda=1/\nu$~\cite{cf}. Hence we expect the value 
$\lambda^{-1}\sim 0.63$ for  the cross-over from the three-dimensional 
to the two-dimensional Ising model in which we are interested.

In view of this analysis the result 
suggested 
in~\cite{cf} (based on the results of a strong 
coupling analysis) that $\lambda^{-1}\sim 0.63$ only for $fixed$ 
bc (in the $N_t$ direction), and that
$\lambda^{-1}\sim 0.5$ for periodic bc seems indeed rather surprising.
It is in fact very unlikely that a 
change in the bc could affect the above analysis, which 
is essentially a bulk result. The only known examples of such a change 
refer to rather degenerate cases like the spherical model
or the crossover from the two-dimensional to the one-dimensional Ising
 model where the choice of the bc can indeed drastically affect the bulk 
of the model, changing for instance its operator content. 

Since our measurements have been made exactly in the case of periodic 
bc we are in the position to test the result of~\cite{cf}. We shall 
give below strong evidence that also for periodic bc 
one has $\lambda^{-1}\sim 0.63$. Most probably the result reported 
in~\cite{cf} was only an artifact of the strong coupling expansion.

\section{The Olesen conjecture}

It is widely accepted that the infrared regime of a LGT can be described 
by some sort of effective string theory. This conjecture is a natural 
consequence of the observation that the colour flux joining a pair of 
quarks in the confining phase  is concentrated inside a tube of  
finite thickness. According to the conjecture, when the quark and the 
antiquark are 
pulled very far apart, the flux tube should behave like a vibrating 
string~\cite{string}. Due to the well known problems related to the Weyl 
anomaly even the simplest possible string theory, the bosonic string, 
cannot be consistently quantized in more than one dimension. Depending 
on the quantization method, one finds either the breaking of rotational 
invariance or the appearance of interacting longitudinal modes 
(Liouville fields).
However it 
is possible to show~\cite{ol2} that for  
large interquark separation these anomalous 
contributions can be neglected. The resulting {\sl effective} string 
theory becomes in this limit 
an ordinary two dimensional conformal field theory. In 
particular, in this infrared limit, the bosonic string flows toward the
ordinary (two-dimensional) free gaussian model. 
In the framework of the lattice 
regularization the interquark potential is given by the expectation 
value of Wilson loops. Thus, according to the above conjecture, we 
expect  the quantum fluctuations of large Wilson loops to be well 
described by some suitable effective string model (not necessarily the 
bosonic one). It can be shown that 
the contribution to the expectation value of the Wilson 
loop due to these quantum fluctuations can be written as
 a power series 
in the adimensional parameter 
$(\sigma A)^{-1}$ 
where $\sigma$ is the ordinary zero temperature string tension and 
$A$ is the Area of the Wilson loop. By using the effective string model 
one can then evaluate (at least in principle) these contributions
 order by order in the expansion parameter. 
This approach was pioneered in~\cite{lsw} (where it was calculated the 
first term of the expansion) and its consequences have been successfully 
tested in various models by several groups during the past years.

This same reasoning can be extended also to the case of the correlator 
of two Polyakov loops: $<P(x)P(x+R)>$. From the string point of view
the only difference with respect to the Wilson 
loop case is in the bc, which can be easily taken into 
account. From the gauge theory point of view we have a new important 
feature: the natural expansion parameter of the theory is now 
$T/\sqrt{\sigma}$. One can thus hope to obtain an universal relation between 
$\sigma$ and $T_c$ by requiring, for instance, that the temperature 
dependent string tension $\sigma(T)$ vanishes at the deconfinement 
temperature $T_c$.

 This program was initiated by O.Alvarez and R.Pisarski~\cite{ap}, 
who worked in the $d\to \infty$ limit (where $d$ are  
the space-time dimensions). Their result was later 
extended by P.Olesen to 
any value of $d$~\cite{ol}. Assuming  the bosonic 
string model as effective description, 
they obtained in the $R\to \infty$ limit
\eq
<P(x)P(x+R)>\sim e^{-\sigma(T)R/T}
\en
where $\sigma(T)$ 
 is given by:
\eq
\sigma(T)=\sigma\sqrt{1-\left(\frac{T}{T_c}\right)^2}
\en
and
\eq
T_c^2/{\sigma}=\frac{3}{\pi(d-2)}.
\label{olesen}
\en
We shall denote in the following eq.(\ref{olesen})
as the ``Olesen conjecture''.

Let us stress again that this result has been obtained assuming the bosonic
string as 
effective model, which could well be a too simple 
description for the infrared regime of realistic lattice gauge theories. 
For instance in eq.(\ref{olesen}) there is no residual information of 
the gauge group of the model. The only remaining dependence is on the 
number of space-time dimensions of the theory. However, recently, 
exactly in the case of the 3d Ising model, we obtained an impressive 
confirmation of this bosonic string picture~\cite{cfghpv}. In order to 
describe this result let us first make  
an important observation (see \cite{cfghpv}  for the details).
The bosonic effective string approach described above
 is completely equivalent to the so called ``capillary wave model''
(CWM)~\cite{cw}, which describes the behaviour of fluid (or rough) 
interfaces in statistical systems. This equivalence can be easily 
understood in the context of the Ising model. The same duality 
transformation which relates the gauge and the spin models, maps Wilson 
loops of the gauge model  into interfaces in the spin model (modulo 
 the obvious changes in the bc), and maps the effective
bosonic string into the CWM. In fact the CWM, in its simplest 
formulation, describes the interface with an effective Hamiltonian 
proportional to its area~\footnote{More precisely, it is proportional to 
the variation of the surface area with respect to the classical 
solution, which in the Ising case is the cristallographic plane.}, and 
this is exactly one of the possible ways to define the bosonic string.
In~\cite{cfghpv} we succeeded
in evaluating the second term of the expansion in $\sigma A$ in the CWM
 and found 
a perfect agreement with the results of a set of high precision 
simulations. If this agreement would be conserved to higher order of the 
expansion it would necessarily imply the correctness of the Olesen 
conjecture. This was one of the initial motivations for the present 
work. Testing the Olesen conjecture we are in some sense testing the 
bosonic string picture to all orders (at least in the Ising model).

\section{The Numerical Method}
\subsection{The model}
We study the spin Ising model defined by eq.(\ref{zspin}) and 
(\ref{hspin}), on a simple cubic lattice with extension $N_x,N_y$ and 
$N_t$ in the three directions $x,y,t$. $N_t$ will be always smaller 
than $N_x$ and $N_y$, and the bc in the $t$ and $x$ 
direction will always be periodic. In the $x$ direction we shall 
implement both periodic and antiperiodic bc. Antiperiodic bc will be 
obtained by setting the couplings $J_{\br ij \kt}=-1$ in the links 
joining the lowermost and uppermost layers in the $x$ direction, while 
keeping all the other couplings equal to $+1$. 
In the following we shall denote the Hamiltonian constructed in this way
as $H(s,ap)$, while the Hamiltonian defined as 
in eq.(\ref{hspin}) with all the $J_{\br ij \kt}=1$
will be denoted by $H(s,p)$. 
In the following an important role will be played by the ratio of 
antiperiodic and periodic 
partition functions $Z_{ap}/Z_{p}$. The reason of this is twofold:
first, as it was shown in~\cite{h93}, it can be used, as the Binder 
cumulant, to estimate the critical temperature of the system and to 
extract the critical index $\nu$. Moreover it was shown in~\cite{h93}, 
that this ratio is  more efficient and converges faster than the 
standard Binder cumulant. Second, in the present case, owing to the 
exact solution of the two dimensional Ising model, and according to
the discussion of 
the previous sections, we can predict exactly the value of the ratio
$Z_{ap}/Z_{p}$ at the critical temperature. This will give us an 
independent, very precise method to 
extract the critical temperature.

In order to obtain the ratio $Z_{ap}/Z_{p}$ one 
must generalize the system and treat also the bc as 
dynamical variables. The partition function of this generalized system 
is given by
\eq
Z=\sum_{bc}\sum_{s_i=\pm1}\exp[-\beta H(s,bc)]
\en
where the new index $bc$ can take the 
 two possible values: $bc=\{p,~ap\}$.
The fraction of configurations with antiperiodic bc is given by:
\eq
\frac{Z_{ap}}{Z}=\frac{\sum_{bc}\sum_{s_i=\pm1}\exp[-\beta H(s,bc)]
\delta_{bc,ap}}{Z}=\br\delta_{bc,ap}\kt
\en
and similarly for the periodic ones.
Thus we can express the ratio $Z_{ap}/Z_{p}$ as a ratio of observables 
in the same system:
\eq
\frac{Z_{ap}}{Z_{p}}=\frac{\br\delta_{bc,ap}\kt}
{\br\delta_{bc,p}\kt},
\en
which is hence accessible in a single Monte Carlo simulation, provided the 
bc can be updated efficiently.
This is the case for the algorithm which shall be described in the 
following subsection.

 \subsection{The algorithm}
For the updating of the bc we used a specially
adapted cluster algorithm
which was introduced in  ref. \cite{fluct}.
Let us shortly recall the basic steps of this algorithm.

First the links of the lattice are deleted with the standard probability
of the Swendsen-Wang \cite{SW} algorithm.
\eq
 p_d = \exp(- \beta (1+J_{<ij>} s_i s_j) ) .
\en
In the second step  one checks whether it is possible to change the type
of bc simultaneously with a sign change of certain spins
without changing the sign of $J_{<ij>} s_i s_j$ for any frozen
(i.e. not deleted) link.

All clusters (in the sense of the standard Swendsen-Wang algorithm) which have
a site in the $x=1$ slice are
checked whether they can be uniquely decomposed into a part "above" and
"below" the $x=1$ to $x=L$ boundary. Technically this is achieved  by
introducing an additional label when constructing the cluster using the
ants in the labyrinth algorithm. The start site of the cluster in the $x=1$
slice gets the label $-1$. The sites joining the cluster get the label of their
parent site, except when the $x=1$ to $x=L$ boundary is crossed. Then 
the new site gets minus the label of the parent site. If at the end of the
construction of the clusters all sites connected by a frozen link have
consistent labels the type of the bc is altered and the spins
are multiplied by their label.

\subsection{The Simulations}
We alternated the boundary flip with standard single cluster updates  \cite{ulli}.
In the case of our largest lattice ($128 \times 128 \times 16$) we performed
32 single cluster updates per boundary flip. After each boundary flip we 
performed a measurement of the energy, the magnetisation and the type of 
the bc. The total number of measurements was 100000 
throughout.  It is interesting to note that the integrated autocorrelation 
time $\tau$ for the type of the bc was smaller then $0.5$ due 
to negative values of the autocorrelation function.  A similar observation was 
already made in~\cite{h93} for the $L_x=L_y=L_t$ case. 

\section{Discussion}
\subsection{Behaviour of $Z_{ap}/Z_{p}$}
The first test of the Svetitsky-Yaffe conjecture is given by the 
behaviour of $Z_{ap}/Z_{p}$. In fact, according to the discussion of 
sect.2 we expect that at the critical point this ratio coincides with 
the corresponding one in the two-dimensional Ising model. This can be 
evaluated by using the exact solution of the model 
 (see for instance~\cite{ff}) or by using well known conformal field 
theory techniques (for a review of this techniques see for 
instance~\cite{id}). It depends only on the shape of the two dimensional 
lattice, namely on the ratio between $N_y$ and $N_x$. The result is:
\eq
\frac{Z_{ap}}{Z_{p}}=\frac{{\vartheta_{2}(0,\tau)}+
{\vartheta_{3}(0,\tau)}-{\vartheta_{4}(0,\tau)}}{
{\vartheta_{2}(0,\tau)}+{\vartheta_{3}(0,\tau)}+{\vartheta_{4}(0,\tau)}}
\label{cft}
\en
where $\tau=iN_y/N_x$, and ${\vartheta_{i}(0,\tau)}$ 
denote the Jacobi theta functions 
(for the notations see for instance ref.~\cite{id}).
For completeness we remind that $Z_{p}$ is periodic both in $x$ and 
$y$ directions, while $Z_{ap}$ is periodic 
in the $y$ direction and {\sl antiperiodic} in the $x$ direction.
Two interesting limits of this equation are the following:
 In the $N_x\to \infty$ limit we get $\frac{Z_{ap}}{Z_{p}}\to1$ as 
expected, since in this limit the bc in the $x$ direction are 
irrelevant. If $N_x=N_y$ (square lattice) we have 
$\frac{Z_{ap}}{Z_{p}}= 0.37288488...$ , a result which we shall use 
in the following.

In order to test this behaviour we fixed $N_t=6$, and first simulated  
the model choosing $N_x=N_y=12,24,48,96$. We extracted the critical 
temperature using the usual crossing procedure (see ref.~\cite{h93}) and 
found the values quoted in tab.1. For each crossing we also evaluated 
the ratio  $Z_{ap}/Z_{p}$ (third column of  tab.1). All the values 
are in good agreement with the predictions of eq.(\ref{cft}). We decided 
thus to {\sl assume} the S-Y conjecture and extract the critical 
temperature by requiring $Z_{ap}/Z_{p}$ to be equal to
 $0.37288488...$. The results are reported in tab.2.
The gain  
in precision with respect to the usual 
crossing method is immediately obvious.
 As our best estimate of the critical 
temperature we take the value
 corresponding to the largest lattice, namely in the 
present case $\beta_c(N_t=6)= 0.228818(4)$. 
Fixing this value of $\beta$ we then studied the system for various 
asymmetric lattices, and extracted the corresponding  $Z_{ap}/Z_{p}$
values. The choices of lattice sizes and the results are reported in 
tab.3  and fig.1 together with the predictions obtained from 
eq.(\ref{cft}). The agreement is impressive in the whole range of 
$\tau$ values. Up to our knowledge this is the first example of such a 
test of the S-Y conjecture and it is probably the most precise and 
compelling evidence of the validity of the conjecture among those that 
we shall present in this paper.

\subsection{The critical temperature $T_c$}

We then moved to other values of $N_t$ keeping the above described 
approach. We always studied  lattices with the same values of $N_x=N_y$. 
We determined the critical temperature $\beta_c(N_t)$ by imposing 
$\frac{Z_{ap}}{Z_{p}}= 0.37288488...$. In tab.4 we report the set of 
lattices that we studied.
 For any given $N_t$ we 
chose as our best estimate of the critical 
temperature that corresponding to the largest value of $N_x$.
The resulting values of $\beta_c$ are listed in the third column of 
tab.4, while in the fourth column the corresponding values 
for the dual gauge model are given, obtained by using eq.(\ref{dual}). 
We shall postpone to
sect.7 a detailed comparison of our results with other existing estimates, but
we have already reported in tab.4 the two sets of estimates:~\cite{wz}
and~\cite{cf} which we discussed in the introduction.
In particular, in the fifth 
column we have reported the Montecarlo 
estimates of~\cite{wz} (for $N_t=2,4,8$) for $\beta_c$.
 They agree with our 
results, which are however more than one order of magnitude more 
precise. On the contrary the estimates obtained by using
 high temperature expansions (last column of tab.4) systematically 
disagree with our results (except for the $N_t=2$ case). 

Let us now study the scaling behaviour of these data.
Based on general renormalization group 
arguments we expect corrections to scaling 
to be of the form
\eq
N_t = c_1 \epsilon^{-\nu} \left(1+ c_2 \epsilon^{\Delta} + c_3 \epsilon ...
\right)
\en
where $\Delta=\nu \omega$, with $\omega \approx 0.82$ \cite{bloete} 
 and $c_1$ is
related to the constant $b_1$ which appears in eq.(\ref{3.2}) by: 
$c_1=(b_1)^{\nu}$. The second 
correction term reflects the ambiguity in the definition of the reduced 
critical point shift
\eq
\epsilon'(N_t) = \frac{T_c(N_t)-T_c}{T_c} =
            \frac{1/\beta_c(N_t)-1/\beta_c}{1/\beta_c}
          = \epsilon(N_t) (1-\epsilon(N_t)...)    
\en
First we tried to perform a fit with $c_1,c_2,c_3,\nu$ and $\Delta$ as 
free parameters, while keeping $\beta_c=0.2216546(10)$ \cite{bloete} fixed.
However even when taking all results summarized in table 4 
the statistical errors of $c_2,c_3$ and $\Delta$ make the result meaningless. 
Therefore we chose a much simpler approach: We solved the equations
\eq
N_t = c_1 \left[\epsilon(N_t)\right]^{-\nu}
\label{eps}
\en
and 
\eq
N_t = c_1 \left[\epsilon'(N_t)\right]^{-\nu}
\label{epsp}
\en
for pairs of $N_t$ with respect to $c_1$ and $\nu$. 
The results are summarized in table 5.  
The results obtained 
from eq.(\ref{eps}) seem to converge faster  than those from eq.(\ref{epsp}).
Therefore we regard $\nu=0.628(3)$ and $1/c_1=1.477(18)$
 obtained from the pair $N_t=12$ and $N_t=16$
and using  eq.(\ref{eps}) as our best approximation.  Comparing the numbers
obtained from eq. (\ref{eps})  and eq. (\ref{epsp}) we still might expect 
systematic errors  of the order of $0.005$ for $\nu$  and $0.05$ for $1/c_1$. 
Our best estimate agrees
perfectly with the existing estimates ($\nu\sim 0.63$) of this critical
index for the 3d Ising model, and disagrees with the estimate
($\nu\sim 0.5$) of ref.~\cite{cf}. 

In addition  this analysis can be performed using the $Z_2$ gauge theory  
inverse temperatures eq.(\ref{dual}) instead of the inverse temperature 
of the Ising spin model. 
\eq
 aT_c = \frac{1}{N_t} \left(\tilde \beta_c -\tilde \beta_c(N_t) \right)^{-\nu} 
\en
 Then we obtain as our best estimates 
$\nu=0.6251(26)$ and $aT_c=2.29(3)$ , to be compared, for instance,
 with the value $aT_c=2.2(1)$ quoted in~\cite{cg91}.

\subsection{Test of the Olesen Conjecture}
In order to obtain an accurate estimate of the ratio $T_c/\sqrt{\sigma}$ 
we need the value of the string tension\footnote{As it is well 
know the string tension of the gauge Ising model and the interface 
tension of the spin Ising model are related by duality and have the same 
value. To simplify the discussion, in the following we shall always refer 
to $\sigma$ as the string tension.} at precisely the critical values of 
$\beta$. A good approximation for these values can be obtained taking as 
reference values those published in~\cite{hp} and~\cite{cfghpv} and 
interpolating among them with the law:
\eq
\sigma(\beta_c)=\sigma(\beta_{ref})\left(\frac{\beta_c-\beta}
{\beta_{ref}-\beta}\right)^{2\nu}
\label{inter}
\en
where $\beta_{ref}$ is the coupling at which the reference value of 
$\sigma$ is taken and $\beta$ is as usual the bulk critical point of the 
3d Ising model. If $|\beta_c-\beta_{ref}|$ is small enough
 then the uncertainty in the 
scaling law and in the determination of $\nu$ can be neglected.
We can evaluate the error for  $\sigma(\beta_c)$ in a reliable way. 
 The results of this analysis are reported in tab. 6.
Once the values of  $\sigma(\beta_c)$ are known one can construct the 
ratio $T_c/\sqrt{\sigma}$ and test if it is stable within the scaling 
region. These ratios are reported in the last column of tab.6. As it can 
be easily seen scaling is well established. Unfortunately this procedure 
cannot be extended up to $N_t=14$ and 16, since in that region $\sigma$ 
has never been estimated.  As our final result for the 
ratio we quote that corresponding to $N_t=12$
\eq
\frac{T_c}{\sqrt{\sigma}}=1.2216(24)~~~.
\label{tcsigma}
\en

To give an idea of the enhancement in precision with respect to the 
previous estimates  we quote the analogous ratio, taken from 
ref.~\cite{cg91}:
${T_c}/{\sqrt{\sigma}}=1.17(10)$.

Our result disagrees from Olesen's prediction
${T_c}/{\sqrt{\sigma}}=\sqrt{3/\pi}=0.977...$ by about 20\%. This is not
too surprising in view of the remarks made in sect.4 . On the contrary it is
rather impressive to see how effective 
 such a simple description can be. It would be very interesting
to see if some improved string model could give better estimates for the
critical temperature.

\subsection{Critical indices}

According to the  Svetitsky-Yaffe conjecture the critical indices $\nu$ 
and $\eta$ which 
characterize the finite temperature transition point at fixed $N_t$
 must coincide with the corresponding indices of the two dimensional 
Ising model, whose values are $\nu=1,~~\eta=0.25$. This is a rather 
important test since in~\cite{wz}, in a set of high precision simulations 
of the 3d Ising model with $N_t=8$, a rather larger value 
for $\eta$ was found: $\eta=0.40(9)$.
 This is indeed the only claim for a failure  
of the S-Y conjecture of which we are aware. 
To extract the two indices we used a standard finite size technique, 
varying the spatial size $N_x$ of the lattice. To obtain the index $\nu$ 
we studied the slope of the $Z_{ap}/Z_{p}$ at the critical point 
which is expected to scale as $N_x^{1/\nu}$. To extract $\eta$ we 
measured the susceptibility $\chi$ which is expected to scale as 
$N_x^{2-\eta}$. To extract the indices we compared nearby values of 
$N_x$. The results are reported in tab.7.
 We expect that the measured values approach the correct asymptotic 
result as $N_x$ increases. It is however interesting to notice that in 
the case of the $\nu$ index this asymptotic value is approached 
precociously, indeed $all$ the values listed in tab.7 are compatible 
within the errors with the expected $\nu=1$ value. This indicates 
that, using the $Z_{ap}/Z_{p}$ slope to extract $\nu$ one has a very
small correction to the scaling amplitude, a fact that was already 
noticed in~\cite{h93}.
On the contrary in the case of the susceptibility we see a large 
correction to scaling and the expected value is in fact reached only for 
the largest lattices that we studied. Smaller lattices give systematically 
higher values of $\eta$. This probably explains the result of 
ref.~\cite{wz}. In conclusion we can say that, also for the critical 
indices,  the S-Y conjecture is 
completely confirmed by our high precision test.

\section{Comparison with existing results}

As we mentioned in the introduction, in these last years the layered Ising model
with low values of $N_t$, in particular with $N_t=2$, have become a
laboratory to test theoretical methods~\cite{cf,ls,acpv,mhz} and numerical
techniques~\cite{wz,kkis,lsz} to estimate the critical indices and the critical
temperature of statistical systems. 
In tab. 8 we have compared our estimates for the critical temperature
with those 
obtained with the mean field transfer matrix approach~\cite{ls}, the
M\"uller-Hartmann-Zittarz method~\cite{mhz}, the constrained variational
approach~\cite{acpv} and strong coupling expansions~\cite{cf}.
Let us make few observation on these numbers (for a more detailed description of
the various approaches we refer to the original literature).
\begin{description}
\item{a]} 
In the transfer matrix version of the mean field
approximation~\cite{ls} one actually deals with clusters of infinite extent. As
a consequence the determination of the critical temperature is greatly improved
with respect to standard mean field results\footnote{In the case of the 2d
Ising model this approach gives indeed the exact result}. This explains the
impressive precision of the result for $N_t=2$, for which the authors
of~\cite{ls} were also able to estimate the theoretical uncertainty. It is very
interesting to observe that our result is in perfect agreement with this
theoretical estimate. As $N_t$ increases the method becomes less reliable,
however in~\cite{ls} the authors give arguments to support the idea that the
$N_t=3$ value is still correct within 4 digits, a result which is again in
perfect agreement with our estimate.  

\item{b]}
 As it can be seen in tab. 8 the 
M\"uller-Hartmann-Zittarz method (MHZ) which was first proposed in~\cite{mhz} 
and studied in this context in~\cite{ls} gives much worse results for the
critical temperature. This is a rather puzzling result
 since MHZ has some features in
common with our approach. MHZ also studies the interface free energy which is
related to the logarithm of the ratio $Z_{ap}/Z_{p}$ that we study. They locate 
the critical point requiring the vanishing of the interface string tension. The
main difference between our approach and MHZ is that they approximate the 
 $Z_{ap}/Z_{p}$ ratio with the partition function of a suitable Solid on Solid
 model (SOS). 
In this way they neglect overhangs and handles of the interface and 
exactly this approximation is the reason of the failure as $N_t$ 
increases (while it
works perfectly in the case of the pure two dimensional model, where it gives
the exact result).
This is indeed very interesting, since it gives a further evidence of the
 relevance of not SOS-like
  configurations in the physics of two dimensional interfaces.
 This was already noticed in~\cite{cgmv}, where the physics of microscopic
 handles in fluid interfaces was discussed, and also in~\cite{cfghpv} where,
in the framework of the Capillary Wave Model (CWM),  the failure of a
simple SOS approximation (which in the CW approach becomes a pure
free bosonic action) for two dimensional interfaces was shown.

\item{c]}
The constrained variational approach~\cite{acpv} has been applied up to now only
to the $N_t=2$ case. The interesting feature of this approach is that it also
allows to extract the critical indices $\nu$ and $\gamma$. In particular
$\gamma$ is obtained by looking at the variation of the critical temperature as
a function of the ratio $\rho$ between the two couplings $\beta_s$ and 
$\beta_t$. In the region in which we are interested, which corresponds to 
$\rho=2$ in the notations of~\cite{acpv} the two indices agree with very high
precision with the values: $\nu=1$, $\gamma=7/4$ (which imply $\eta=1/4$)
that we have found and that are expected from the SY conjecture.

\end{description}

As $N_t$ increases all the theoretical approaches start to give diverging
results. For these values of $N_t$ the only reliable tool to estimate the
critical temperature are the Montecarlo simulations. The existing results
are reported in tab.9 and compared with our ones. All the estimates quoted in
tab.9  except our ones have been obtained by using the method of
Binder cumulants. As it can be seen all the estimates agree perfectly 
within the errors. Notice however that our results are in general one order of
magnitude more precise than the others. This seems to indicate that the
method that we have used in this work (which was originally 
proposed in~\cite{h93}) is indeed superior to the standard Binder cumulants 
method.

\vskip 1cm
{\bf  Acknowledgements}

We thank F.Gliozzi, K.Pinn, P.Provero and S.Vinti for many helpful 
discussions. 
M. Hasenbusch expresses his gratitude
for support by the Leverhulme Trust under grant
16634-AOZ-R8 and by PPARC.

\newpage
{\Large{\bf Tables}}
\vskip 2cm.

\vskip0.3cm
$$\vbox {\offinterlineskip
\halign  { \strut#& \vrule# \tabskip=.5cm plus1cm
& \hfil#\hfil
& \vrule# & \hfil# \hfil
& \vrule# & \hfil# \hfil &\vrule# \tabskip=0pt \cr \noalign {\hrule}
&& $N_x$ && $\beta_c$ && $Z_{a.p.}/Z_{p.}$ &
\cr \noalign {\hrule}
&& $12-24$ && $0.22878(5)$ && $0.376(4)$  & \cr \noalign {\hrule}
&& $24-48$ && $0.22881(3)$ && $0.373(4)$  & \cr \noalign {\hrule}
&& $48-96$ && $0.22882(1)$ && $0.370(4)$  & \cr \noalign {\hrule}
}}$$
\begin{center}
{\bf Tab. 1}{\it~~ Results of the crossing analysis for $N_t=6$.
In the first column the two values of $N_x$ used to extract $\beta_c$, 
which is reported in the second column.
In the last column the value of 
 $Z_{a.p.}/Z_{p.}$ at the crossing point.}
\end {center}
\vskip3cm

\vskip0.3cm
$$\vbox {\offinterlineskip
\halign  { \strut#& \vrule# \tabskip=.5cm plus1cm
& \hfil#\hfil
& \vrule# & \hfil# \hfil
& \vrule# & \hfil# \hfil &\vrule# \tabskip=0pt \cr \noalign {\hrule}
&& $N_x$ && $\beta_c$ && $Slope$ &
\cr \noalign {\hrule}
&& $12$ && $0.22884(4)$  && $-54.4(4)$ & \cr \noalign {\hrule}
&& $24$ && $0.22881(2)$   && $-107.9(7)$ & \cr \noalign {\hrule}
&& $48$ && $0.228812(9)$  && $-219.6(1.9)$ & \cr \noalign {\hrule}
&& $96$ && $0.228818(4)$ && $-434.1(4.3)$  & \cr \noalign {\hrule}
}}$$
\begin{center}
{\bf Tab. 2}{\it~~ $\beta_c$ values for $N_t=6$ and various choices of 
$N_x$, obtained by imposing
$Z_{a.p.}/{Z_{p.}}= 0.37288488...$. In the last column is reported 
the slope of $Z_{a.p.}/{Z_{p.}}$ at the critical point.}
\end {center}
\vskip3cm

\newpage

\vskip0.3cm
$$\vbox {\offinterlineskip
\halign  { \strut#& \vrule# \tabskip=.5cm plus1cm
& \hfil#\hfil
& \vrule# & \hfil# \hfil
& \vrule# & \hfil# \hfil
& \vrule# & \hfil# \hfil &\vrule# \tabskip=0pt \cr \noalign {\hrule}
&& $N_x$ && $N_y$ && $Z_{a.p.}/Z_{p.}$ && $eq.(\ref{cft})$ &
\cr \noalign {\hrule}
&& $32$ && $96$ && $0.0858(11)(4)$  && $0.0867$ & \cr \noalign {\hrule}
&& $48$ && $96$ && $0.1778(15)(8)$  && $0.1752$ & \cr \noalign {\hrule}
&& $64$ && $96$ && $0.2461(18)(12)$  && $0.2491$ & \cr \noalign {\hrule}
&& $96$ && $64$ && $0.5279(23)(15)$  && $0.5292$ & \cr \noalign {\hrule}
&& $96$ && $48$ && $0.6531(20)(12)$  && $0.6558$ & \cr \noalign {\hrule}
&& $96$ && $32$ && $0.8248(20)(7)$  && $0.8269$ & \cr \noalign {\hrule}
}}$$
\begin{center}
{\bf Tab. 3}{\it~~  $Z_{a.p.}/Z_{p.}$ values as a function of 
the lattice shape. In the third column the values measured in the MC 
simulation, the first error is the statistical error, the second takes 
into account the statistical error in the estimate of 
$\beta_c=0.228818(4)$. In the last column the values of eq.(\ref{cft}).}
\end {center}
\vskip3cm

\vskip0.3cm
$$
\hskip-2cm
\vbox {\offinterlineskip
\halign  { \strut#& \vrule# \tabskip=.5cm plus1cm
& \hfil#\hfil
& \vrule# & \hfil# \hfil
& \vrule# & \hfil# \hfil
& \vrule# & \hfil# \hfil
& \vrule# & \hfil# \hfil
& \vrule# & \hfil# \hfil &\vrule# \tabskip=0pt \cr \noalign {\hrule}
&& $N_t$ && $N_x$ && $\beta_c$ && $\tilde{\beta_c}$ && $\cite{wz}$
 && $\cite{cf}$ & \cr \noalign {\hrule}
&& $2$ && $4,8,16,32$ && $0.27604(3)$  && $0.65608(5)$ && $0.2758(4)$
&& $0.2760(10)$  & \cr \noalign {\hrule}
&& $3$ && $24$ && $0.24607(8)$  && $0.71102(8)$ && 
&& $0.2449(1)$  & \cr \noalign {\hrule}
&& $4$ && $8,16,32,64$ && $0.236025(8)$  && $0.73107(2)$ && $0.2358(2)$
&& $0.2347(1)$  & \cr \noalign {\hrule}
&& $5$ && $40$ && $0.231421(13)$  && $0.74057(3)$ && 
&& $0.2301(2)$  & \cr \noalign {\hrule}
&& $6$ && $12,24,48,96$ && $0.228818(4)$  && $0.746035(8)$ && 
&& $0.2273(3)$  & \cr \noalign {\hrule}
&& $7$ && $52$ && $0.227195(9)$  && $0.74947(2)$ && 
&& $0.2258(3)$  & \cr \noalign {\hrule}
&& $8$ && $16,32,64$ && $0.226102(5)$  && $0.75180(1)$ && $0.2262(2)$
&& $0.2248(1)$  & \cr \noalign {\hrule}
&& $10$ && $80$ && $0.224743(5)$  && $0.75472(1)$ && 
&&   & \cr \noalign {\hrule}
&& $12$ && $24,48,96$ && $0.223951(3)$  && $0.756427(6)$ && 
&&   & \cr \noalign {\hrule}
&& $14$ && $102$ && $0.223442(4)$  && $0.757527(8)$ && 
&&   & \cr \noalign {\hrule}
&& $16$ && $32,64,128$ && $0.223101(2)$  && $0.758266(4)$ &&
&&   & \cr \noalign {\hrule}
}}$$
\begin{center}
{\bf Tab. 4}{\it~~  $\beta_c$ values for various $N_t$. In the first 
column the inverse temperature $N_t$, in the second column the set of 
lattices sizes which have been simulated, in the third column our best 
estimates for $\beta_c(N_t)$ and in the fourth column the corresponding 
dual values  $\tilde\beta_c(N_t)$. In the last two column are reported 
for comparison some existing data for $\beta_c(N_t)$ taken from 
ref~\cite{wz} (Montecarlo simulations) and 
ref~\cite{cf} (strong coupling expansion).}
\end {center}
\vskip3cm

\vskip0.3cm
$$
\vbox {\offinterlineskip
\halign  { \strut#& \vrule# \tabskip=.5cm plus1cm
& \hfil#\hfil
& \vrule# & \hfil# \hfil
& \vrule# & \hfil# \hfil
& \vrule# & \hfil# \hfil
& \vrule# & \hfil# \hfil &\vrule# \tabskip=0pt \cr \noalign {\hrule}
&&$N_t$&& $1/c_1$   &&   $\nu$ &&   $1/c_1$ &&  $\nu$   & \cr \noalign {\hrule}
&&$2-4$  && 1.3043(10)&&0.5903(3)&&1.0393(7)&&0.5208(3) & \cr \noalign {\hrule}
&&$4-6$  && 1.377(3) &&0.6096(7) &&1.230(3) &&0.5824(7) & \cr \noalign {\hrule}
&&$5-7$  && 1.394(11)&&0.6135(23)&&1.276(9) &&0.5935(22)& \cr \noalign {\hrule}
&&$6-8$  && 1.423(8) &&0.6190(16)&&1.323(8) &&0.6035(16)& \cr \noalign {\hrule}
&&$7-10$ && 1.438(14)&&0.6219(24)&&1.357(13)&&0.6103(24)& \cr \noalign {\hrule}
&&$8-12$ && 1.442(10)&&0.6224(16)&&1.375(9) &&0.6134(16)& \cr \noalign {\hrule}
&&$10-14$&& 1.438(20)&&0.6218(32)&&1.386(19)&&0.6152(31)& \cr \noalign {\hrule} 
&&$12-16$&& 1.476(18)&&0.6275(26)&&1.432(18)&&0.6223(26)& \cr \noalign {\hrule}
}}$$

\begin{center}
{\bf Tab. 5}{\it~~ The critical index $\nu$ and the constant $1/c_1$ as a
function of the pair of $N_t$ values listed in the first column. In the second
and third column the results obtained by using eq.(\ref{eps}). In the last two
columns those obtained by using eq.(\ref{epsp}).}
\end {center}
\vskip3cm

\vskip0.3cm
$$
\vbox {\offinterlineskip
\halign  { \strut#& \vrule# \tabskip=.5cm plus1cm
& \hfil#\hfil
& \vrule# & \hfil# \hfil
& \vrule# & \hfil# \hfil
& \vrule# & \hfil# \hfil &\vrule# \tabskip=0pt \cr \noalign {\hrule}
&& $N_t$ && $\beta_c$ && $a^2\sigma$ && $T_c/\sqrt{\sigma}$ 
& \cr \noalign {\hrule}
&& $6$ && $0.228818(4)$ && $0.0189(2)$  && $1.212(6)$ 
& \cr \noalign {\hrule}
&& $7$ && $0.227195(9)$ && $0.0138(2)$  && $1.216(9)$ 
& \cr \noalign {\hrule}
&& $8$ && $0.226102(5)$ && $0.0105(2)$  && $1.220(12)$ 
& \cr \noalign {\hrule}
&& $10$ && $0.224743(5)$ && $0.00662(12)$  && $1.228(11)$ 
& \cr \noalign {\hrule}
&& $12$ && $0.223951(3)$ && $0.004654(18)$  && $1.2216(24)$ 
& \cr \noalign {\hrule}
}}$$
\begin{center}
{\bf Tab. 6}{\it~~The adimensional ratio $T_c/\sqrt{\sigma}$ for various 
values of $N_t$. In the second column the critical couplings, in the 
third column the corresponding values of $\sigma$ interpolated according 
to eq.(\ref{inter}). The reference values of $\sigma$ for the 
interpolation are taken from  tab.2 of ref.~\cite{hp} for $N_t=6,7,8$ and
$10$, while for $N_t=12$ we have used the value for $\beta=0.224$ 
reported in~\cite{cfghpv}.}
\end {center}
\vskip3cm

\vskip0.3cm
$$
\vbox {\offinterlineskip
\halign  { \strut#& \vrule# \tabskip=.5cm plus1cm
& \hfil#\hfil
& \vrule# & \hfil# \hfil
& \vrule# & \hfil# \hfil
& \vrule# & \hfil# \hfil &\vrule# \tabskip=0pt \cr \noalign {\hrule}
&& $N_t$ && $N_x$ && $\nu$ && $\eta$ 
& \cr \noalign {\hrule}
&& $2$ && $8-4$ && $0.979(21)$  && $$ & \cr \noalign {\hrule}
&& $2$ && $16-8$ && $0.980(21)$  && $$ & \cr \noalign {\hrule}
&& $2$ && $32-16$ && $1.012(21)$  && $$ & \cr \noalign {\hrule}
&& $4$ && $16-8$ && $0.971(25)$  && $$ & \cr \noalign {\hrule}
&& $4$ && $32-16$ && $0.996(21)$  && $$ & \cr \noalign {\hrule}
&& $4$ && $64-32$ && $1.000(23)$  && $$ & \cr \noalign {\hrule}
&& $6$ && $24-12$ && $1.012(20)$  && $0.279(6)$ & \cr \noalign {\hrule}
&& $6$ && $48-24$ && $0.975(21)$  && $0.262(6)$ & \cr \noalign {\hrule}
&& $6$ && $96-48$ && $1.017(28)$  && $0.251(5)$ & \cr \noalign {\hrule}
&& $8$ && $32-16$ && $0.979(20)$  && $$ & \cr \noalign {\hrule}
&& $8$ && $64-32$ && $0.983(21)$  && $$ & \cr \noalign {\hrule}
&& $12$ && $48-24$ && $0.994(24)$  && $$ & \cr \noalign {\hrule}
&& $12$ && $96-48$ && $0.983(26)$  && $0.263(7)$ & \cr \noalign {\hrule}
&& $16$ && $64-32$ && $0.989(19)$  && $0.276(8)$ & \cr \noalign {\hrule}
&& $16$ && $128-64$ && $1.026(26)$  && $0.263(7)$ & \cr \noalign {\hrule}
}}$$
\begin{center}
{\bf Tab. 7}{\it~~The $\nu$ and $\eta$ indices for various values of 
$N_t$. In the second column the pair of $N_x$ values used to extract the 
index (see text). The susceptibility has been evaluated only in a 
subset of cases.}
\end {center}

\vskip0.3cm
$$
\vbox {\offinterlineskip
\halign  { \strut#& \vrule# \tabskip=.5cm plus1cm
& \hfil#\hfil
& \vrule# & \hfil# \hfil
& \vrule# & \hfil# \hfil
& \vrule# & \hfil# \hfil
& \vrule# & \hfil# \hfil
& \vrule# & \hfil# \hfil &\vrule# \tabskip=0pt \cr \noalign {\hrule}
&& $N_t$ && $\beta_c$ && $\cite{ls}$ && $\cite{mhz}$ && $\cite{acpv}$
 && $\cite{cf}$ & \cr \noalign {\hrule}
&& $2$  && $0.27604(3)$  && $0.276030(2)$ && $0.2847$ && $0.2759$
&& $0.2760(10)$  & \cr \noalign {\hrule}
&& $3$  && $0.24607(8)$  && $0.2460$ && $0.2626$ &&
&& $0.2449(1)$  & \cr \noalign {\hrule}
&& $4$  && $0.236025(8)$  && $0.2390$ && $0.2567$ &&
&& $0.2347(1)$  & \cr \noalign {\hrule}
&& $5$  && $0.231421(13)$  && $0.2357$ && &&
&& $0.2301(2)$  & \cr \noalign {\hrule}
&& $6$  && $0.228818(4)$  &&  && &&
&& $0.2273(3)$  & \cr \noalign {\hrule}
&& $7$  && $0.227195(9)$  &&  && &&
&& $0.2258(3)$  & \cr \noalign {\hrule}
&& $8$  && $0.226102(5)$  &&  && &&
&& $0.2248(1)$  & \cr \noalign {\hrule}
}}$$
\begin{center}
{\bf Tab. 8}{\it~~ Comparison of  $\beta_c$ values obtained with 
different theoretical approaches for small values of $N_t$. In the first 
column the inverse temperature $N_t$ and in the second column  our best 
estimates for $\beta_c(N_t)$. In the following columns the $\beta_c$ values
obtained with the mean field transfer matrix approach~\cite{ls}, the
M\"uller-Hartmann-Zittarz method~\cite{mhz}, the constrained variational
approach~\cite{acpv} and strong coupling expansions~\cite{cf}.}
\end {center}
\vskip3cm

\vskip0.3cm
$$
\vbox {\offinterlineskip
\halign  { \strut#& \vrule# \tabskip=.5cm plus1cm
& \hfil#\hfil
& \vrule# & \hfil# \hfil
& \vrule# & \hfil# \hfil
& \vrule# & \hfil# \hfil
& \vrule# & \hfil# \hfil &\vrule# \tabskip=0pt \cr \noalign {\hrule}
&& $N_t$ && $\beta_c$ && $\cite{kkis}$ && $\cite{wz}$ && $\cite{lsz}$
& \cr \noalign {\hrule}
&& $2$  && $0.27604(3)$  && && $0.2758(4)$ && $0.27598(1)$
& \cr \noalign {\hrule}
&& $3$  && $0.24607(8)$  && $0.24600(8)$ &&  &&
 & \cr \noalign {\hrule}
&& $4$  && $0.236025(8)$  && $0.23605(5)$ && $0.2358(2)$ &&
 & \cr \noalign {\hrule}
&& $5$  && $0.231421(13)$  && $0.23137(5)$ && &&
 & \cr \noalign {\hrule}
&& $6$  && $0.228818(4)$  && $0.22882(3)$ && &&
 & \cr \noalign {\hrule}
&& $7$  && $0.227195(9)$  && $0.22721(4)$ && &&
 & \cr \noalign {\hrule}
&& $8$  && $0.226102(5)$  &&  && $0.2262(2)$ &&
 & \cr \noalign {\hrule}
}}$$
\begin{center}
{\bf Tab. 9}{\it~~ Comparison of  $\beta_c$ values obtained with Montecarlo
simulations by different groups. In the first 
column the inverse temperature $N_t$ and in the second column  our best 
estimates for $\beta_c(N_t)$. In the following columns the $\beta_c$ values
obtained in~\cite{kkis},~\cite{wz} and~\cite{lsz} respectively.}
\end {center}
\vskip 3cm
{\Large{\bf Figure Caption}}
\vskip 0.5cm.

\begin{center}
{\bf Fig. 1}{\it~~ $Z_{ap}/Z_{p}$ values as a function of the ratio 
$N_y/N_x$ for $N_t=6$, at the critical coupling $\beta_c=0.228818$. The crosses
denote the numerical estimates reported in Tab. 3, while the solid line is 
the theoretical expectation according to eq.(\ref{cft}). Errors are not
reported since they are smaller than the symbols.}
\end {center}
\vskip 2cm.

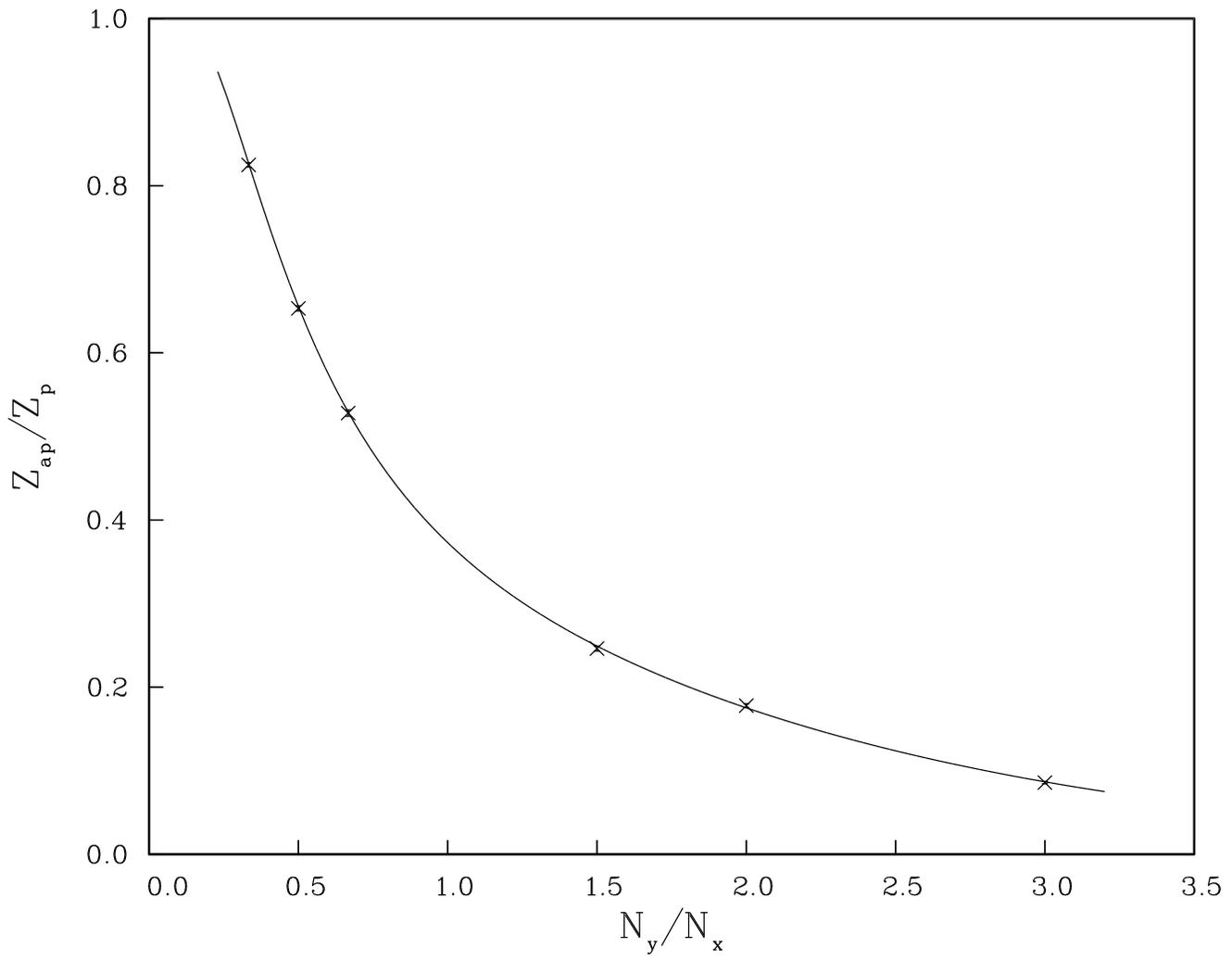
\end{document}